\documentclass[12pt]{article}

\usepackage{newtxtext,newtxmath}

\usepackage{graphicx}
\usepackage{hyperref}

\usepackage[letterpaper,margin=1in]{geometry}

\renewenvironment{abstract}
	{\quotation}
	{\endquotation}

\date{}

\makeatletter
\renewcommand{\fnum@figure}{\textbf{Figure \thefigure}}
\renewcommand{\fnum@table}{\textbf{Table \thetable}}
\makeatother

\usepackage{url}

\title{Dictionary-Based Reconstruction of Spatio-Temporal 3D Magnetic Field Images from Quantum Diamond Microscope}

\author{
	Anuj Bathla$^{1,2}$,
	Madhur Parashar$^{2}$,
	Matthew Markham$^{3}$,
    Ajit Rajwade$^{4}$,\and
    Kasturi Saha$^{2\ast}$
	\\
    \small\parbox[t]{\linewidth}{\centering $^{1}$Center for Research in Nanotechnology and Science,\\ Indian Institute of Technology Bombay, Mumbai 400076, India.}\and
    \small\parbox[t]{\linewidth}{\centering $^{2}$Department of Electrical Engineering,\\ Indian Institute of Technology Bombay, Mumbai 400076, India.}\and
    \small$^{3}$Element Six Co., Didcot, Oxfordshire, OX11 0QR, UK.\and
    \small\parbox[t]{\linewidth}{\centering $^{4}$Department of Computer Science and Engineering,\\ Indian Institute of Technology Bombay, Mumbai 400076, India.}\and
	\small$^\ast$Corresponding author: kasturis@iitb.ac.in\and
}

\begin{document}

\maketitle

\begin{abstract} \bfseries \boldmath
Three-dimensional magnetic imaging with high spatio-temporal resolution is critical for
probing current paths in various systems, from biosensing to microelectronics. Conventional 2D Fourier-based current source localization methods are ill-posed in multilayer or dynamic systems due to signal overlap and noise. In this work, we demonstrate an innovative nitrogen-vacancy (NV) center-based wide-field magnetic microscopy technique for dynamic three-dimensional imaging and localization of current sources. Using custom-fabricated multilayer micro-coil platform to emulate localized, time-varying currents similar to neuronal activity, we acquire magnetic field maps with micrometre-scale spatial and millisecond-scale temporal resolution using per-pixel lock-in-based detection. Source localization and image reconstruction are achieved using a Least Absolute Shrinkage and Selection Operator (LASSO)-based reconstruction framework that incorporates experimentally measured basis maps as spatial priors. Our method enables robust identification of active current sources across space and time, and significantly advances the accuracy of dynamic 3D current imaging and NV-based magnetometry for complex systems.
\end{abstract}

Over the past decade, quantum diamond microscopy, utilizing nitrogen-vacancy (NV) defect centers in diamond, has emerged as a powerful tool for imaging two-dimensional (2D) current density profiles across various samples\,\cite{tetienne2017quantum,turner2020magnetic,kehayias2022measurement}. This technique has been applied to a broad range of systems, from biological applications to condensed matter physics and semiconductor devices\, \cite{hall2012high,sage2013optical,schirhagl2014nitrogen,glenn2015single,barry2016optical,price2020widefield,ariyaratne2018nanoscale,lillie2019imaging,ku2020imaging,scholten2022imaging,Li2022PCB,zhong2024high}. The unique advantages of NV-based magnetic microscopy—such as ambient temperature operation, sub-micron spatial resolution, and the ability to dynamically decouple high-frequency magnetic fields—make it a superior alternative to conventional magnetometers like superconducting quantum interference devices (SQUIDs) and rubidium-vapor magnetometers for imaging based applications\,\cite{taylor2008high,steinert2010high,pham2011magnetic,doherty2013nitrogen,steinert2013magnetic,rondin2014magnetometry,knauss2001scanning,PaolaHighFreqPhysRevX2022,braje2024HighFreq}.

Quantum Diamond microscopy employs a thin, high-density NV layer whose fluorescence is mapped onto a charge-coupled device (CCD) or complementary metal-oxide-semiconductor (CMOS) camera, enabling the extraction of microscopic 2D magnetic field maps through non-linear fitting of the optically detected magnetic resonance (ODMR) spectrum of NV centers\,\cite{levine2019principles,scholten2021widefield}. Further, static microscopy can be extended to temporal maps using lock-in based detection techniques \cite{TemporalSciRep2022, Ulrich2022Lockin}. In numerous studies—especially in condensed matter physics \cite{casola2018probing,shishir2021} and non-invasive imaging of semiconductor chips \cite{scholten2022imaging}—accurately transforming magnetic field maps into current density distributions is essential. While direct Fourier-based inversion techniques have been widely used for 2D current reconstruction in planar systems\,\cite{roth1989using}, their applicability is limited in multilayer geometries or dynamic scenarios due to sensitivity to noise, boundary artifacts, and assumptions of spatial homogeneity \cite{barry2016optical}. Recent advancements have focused on enhancing magnetic field mapping for current density reconstruction by incorporating artificial intelligence, such as MAGIC-UNet, a deep convolutional neural network that significantly outperforms the Fourier method, particularly in the presence of high noise or large standoff distances \cite{ReedMagicNet2025,DNNZhang2024}. Additionally, Bayesian techniques have been explored for current density reconstruction\,\cite{Siddhant2024PRapp}. Recent studies have also reported demonstrations of three-dimensional static current density mapping for electronic circuits via reconstruction of lateral current densities ($J_{xy}$) from magnetic field components ($B_x, B_y$) by numerically inverting the Biot–Savart law \cite{GarsiPRApp2024}. In this work, the authors use linecut analysis and infinite-wire approximations to resolve current contributions from different stacked layers within multilayer devices, identifying failures localized to specific layers. However, several key limitations persist: true 3D magnetic field and current reconstruction is compromised by signal overlap from different layers; reconstruction in thick devices requires prior layout information; spatial resolution is limited by device structure; the 3D inverse problem remains under-determined\,\cite{Utsugi2019}; and mapping magnetic field in 3D with vertical currents ($J_z$) necessitates more advanced algorithms. Furthermore, the approaches employed in earlier NV studies are poorly suited for real-time or dynamic three-dimensional (3D) reconstructions, particularly in applications like neuronal imaging. These methods depend on assumptions of homogeneity and lack the adaptability to integrate temporal dynamics or spatial priors, limiting their effectiveness for complex or rapidly changing current distributions. To address some of these limitations, various regularized inversion methods have been proposed. Tikhonov regularization\,\cite{kolehmainen1994tikhonov} and pseudo-inverse matrix inversion suppress noise by penalizing large solution norms, typically through L2 regularization. However, the use of L2 norms inherently favors smooth solutions and distributes signal energy broadly, which leads to over-smoothing and poor spatial resolution of localized or sparse current features. This limitation is well-established in sparse signal recovery, where L2 minimization tends to spread the solution energy across many basis functions, thereby failing to recover sparse or spatially confined sources\,\cite{chen2001atomic}. Consequently, these methods often struggle in complex scenarios, such as multilayer geometries or densely packed configurations, where overlapping fields from adjacent current-carrying elements are difficult to disentangle\,\cite{nowodzinski2015nitrogen}. More recent inverse approaches—such as time-resolved field-to-current inversion using dynamic basis functions in chemical sensing—offer improved temporal resolution\,\cite{Acheson2023}. However, such methods require complex calibration procedures, are computationally intensive, and often lack scalability for real-time, large-scale imaging.\\
To overcome these challenges, we develop and implement a LASSO (Least Absolute Shrinkage and Selection Operator)-based reconstruction framework tailored to the unique demands of NV widefield magnetometry for applications in spatio-temporally varying 3D imaging. Departing from traditional Fourier-based and linear inversion techniques, this approach leverages the sparsity and structure inherent in current distributions, where typically only a subset of sources is active at a given time—such as in neuronal spiking or localized microelectronic faults. LASSO introduces L1 regularization to promote sparsity, selectively activating the most relevant basis maps while suppressing noise and background correlations. This choice is further motivated by the structure of our forward model, where ill-conditioning and near-singular behavior limit the effectiveness of direct inversion; under such conditions, LASSO offers stable and interpretable solutions.  Unlike previous methods, our approach incorporates structural priors from experimentally acquired basis maps and leverages temporal correlations in the data, allowing accurate recovery of static and dynamic current sources.\\
To validate our framework, we fabricate a novel three-layer micro-coil sample with $\ 40 \,\mu\text{m}$-sized arrays of micro-coils (with $\ 2 \,\mu\text{m}$-wide current tracks) separated by insulating layers. These micro-coils emulate localized and dynamic magnetic field features, providing a testbed for evaluating reconstruction algorithms. By controlling the driving currents, we establish a ground-truth dataset for assessing reconstruction performance under well-defined and tunable conditions.\\
Our results demonstrate that the LASSO-based framework enables highly accurate static reconstruction and meaningful temporally resolved reconstruction of current sources. In static scenarios, across diverse activation patterns, reconstructions achieve strong structural similarity (SSIM: 0.57–0.82), low relative root-mean-square error (RRMSE: 0.20–0.29), and Pearson correlation coefficients exceeding 0.95 with ground-truth magnetic field maps—highlighting the method’s ability to resolve complex, overlapping current sources with high fidelity. For dynamic reconstructions of a 300 ms event involving temporally varying current inputs, we assess performance across different sparsity levels. Temporal reconstructions show moderate accuracy—with SSIM values up to 0.31, RRMSEs ranging from 0.36 to 0.54, and correlation values between 0.29 and 0.76—preserving meaningful spatial and temporal structure despite the presence of noise and overlapping source activity. These results underline the robustness of our approach in static regimes and its promise for further refinement in dynamic applications.\\
Our method demonstrates millisecond-scale temporal resolution, making it ideal for real-time current tracking. This work can be potentially generalized across varying spiking conditions, micro-coil densities, and activation patterns. These capabilities enable not only static mapping but also real-time tracking of transient current dynamics.\\
Overall, our results show that LASSO-based reconstruction provides noise-resilient source estimation, enhancing the capabilities of NV-based magnetic microscopy for imaging complex current distributions. This has significant implications for neuroscience—enabling label-free imaging of neural and cardiac activity—and for semiconductor diagnostics, where it offers a powerful tool for non-invasive analysis of high-density interconnects and transient faults. By bridging the gap between biological and semiconductor imaging, our work advances the field of NV-based magnetometry and opens new avenues for high-resolution, real-time current mapping in diverse applications.

\begin{figure}[hbt!] 
	\centering
	\includegraphics[width=\textwidth]{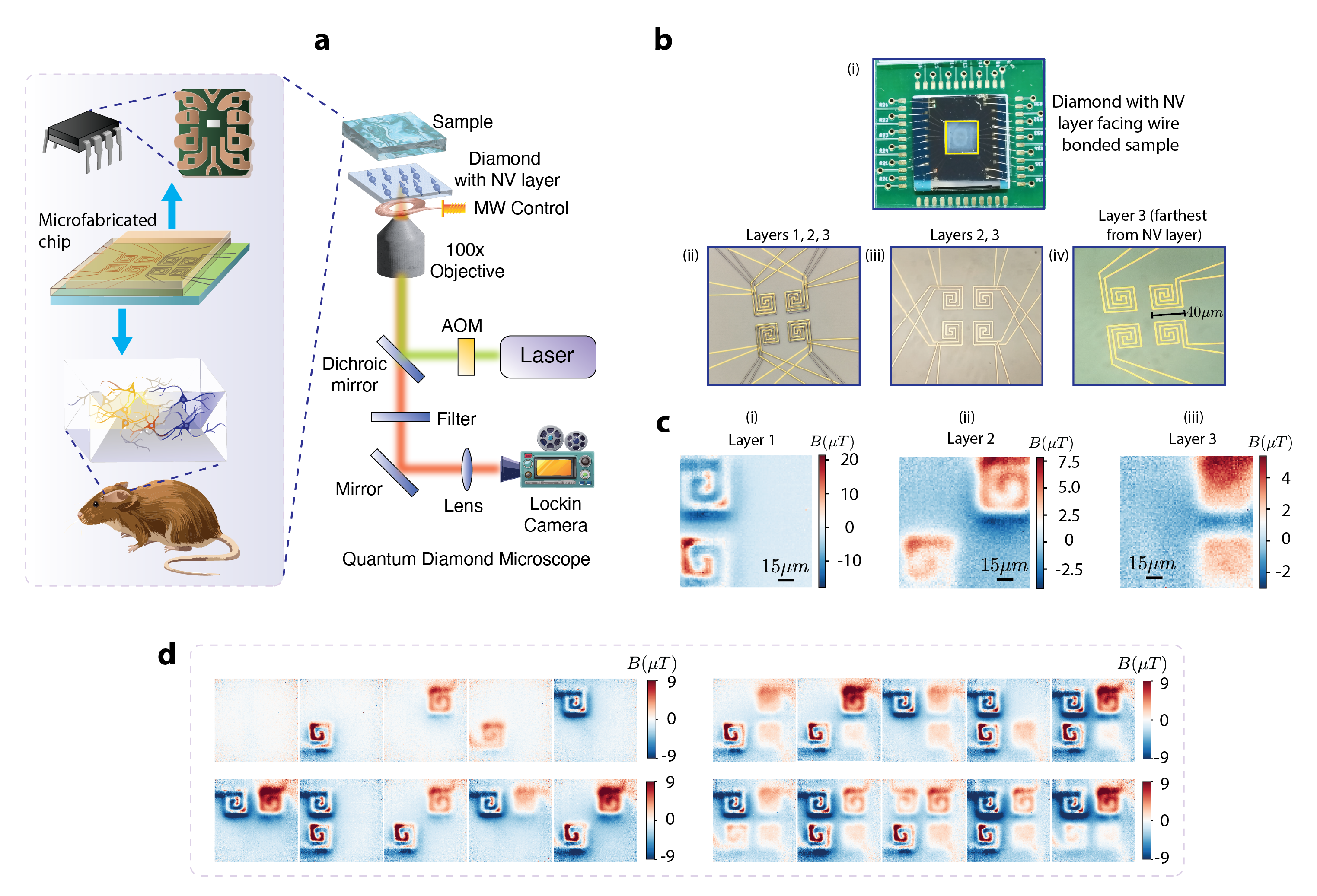}
	\caption{\textbf{Widefield quantum diamond microscope and static magnetic field imaging of 3D micro-coil array.}
		(\textbf{a}) Schematic of the quantum diamond microscope for imaging magnetic fields from fabricated 3D micro-coil array designed to emulate neuronal spiking activity or microchip.(\textbf{b(i)}) Photograph of printed circuit board with micro-fabricated sample mounted with diamond NV center layer facing the circuits. (\textbf{b(i)-(iv)}) Microscope image of micro-coils fabricated on Si substrate. Layer 1 is the closest to the diamond NV layer with least standoff. Layers 1, 2 and 3 are fabricated with insulating layers between them. (\textbf{c(i)- (iii)}) Magnetic field images of two randomly chosen micro-coils excited in individual layers showing out-of-plane field component of B-field measured along one of the NV axis. (\textbf{d}) Subset of static magnetic field maps acquired from 256 combinatorially activated coil configurations.}
	\label{fig1} 
\end{figure}

\subsection*{Results}

\paragraph{Static Magnetic Field Imaging and Reconstruction} \mbox{}\\
NV-based wide-field magnetic imaging provides a powerful approach for capturing static current distributions in micro-coil arrays with high spatial resolution. Static magnetic field imaging of the $3$D multi-layer micro-coil sample was performed using a widefield lock-in NV diamond microscope shown in Figure~\ref{fig1}(a). The sample comprises  of $12$ micro-coils distributed across three layers, with four micro-coils in each layer. Electrical insulation between the layers was achieved using a $\sim 4 \mu m$-thick SU-8 resist polymer, which provides excellent mechanical stability, uniform thickness, and compatibility with standard lithographic techniques. Unlike vapor-deposited dielectric materials, SU-8 enables precise, planar insulation without the need for complex deposition equipment. The micro-coils, fabricated from $100 nm$ thick Ti/Au, each with a size of $40 \, \mu\text{m}$ are driven via $25 \mu m$ wire bonds connected to a custom-made current control printed circuit board (PCB). However, due to microscopic discontinuities and erosion in the $\ 2\mu\text {m}$ track widths, four of the twelve micro-coils became non-conductive, resulting in eight functional coils for imaging and analysis. Further details on the sample and fabrication process are provided in the Methods and Supplementary Section 1.\\
We investigated $256$ distinct magnetic field configurations, corresponding to all  $(2^8)$  possible on/off combinations of the eight working micro-coils across the three layers. Each coil was driven with a DC current of approximately $\sim 600 \,\mu\text{A}$, and magnetic field images were acquired using lock-in detection at a $6.25 kHz$ microwave modulation frequency. The magnetic field for each configuration was projected along a single NV axis and obtained by sampling the NV resonance spectrum with a frequency step size of $100 kHz$. The ODMR corresponding to each pixel in the lock-in camera was non-linearly fit to extract the local resonance frequency. The magnetic field distribution was determined from the resonance frequency shifts between the current on and off states, referenced against a background bias field (see Supplementary Section 2).
Enabled by high-speed lock-in imaging, the full dataset was acquired within ~12 hours, with each configuration captured in under 3 minutes—significantly faster than conventional CMOS-based NV magnetic microscopy approaches that typically require several minutes per scan \cite{davis2018mapping}.\\
A subset of representative field maps is shown in Figure ~\ref{fig1}(d), with a spatial resolution of $\sim 1.2 \,\mu\text{m}$ sufficient to resolve current tracks as narrow as $2 \,\mu\text{m}$. Peak magnetic field amplitudes across all pixels were generally below $10\,\mu\text{T}$. Due to a thin layer of UV-cured optically transparent adhesive, the closest micro-coils layer is at $\sim 7 \,\mu\text{m}$ from the NV surface, while the other layers are further away, resulting in increasing spatial blurring in deeper layers due to greater standoff. This depth-dependent resolution degradation is especially relevant for applications such as integrated circuit imaging and biological current mapping.

\begin{figure}[hbt!]
	\centering
	\includegraphics[width=\textwidth]{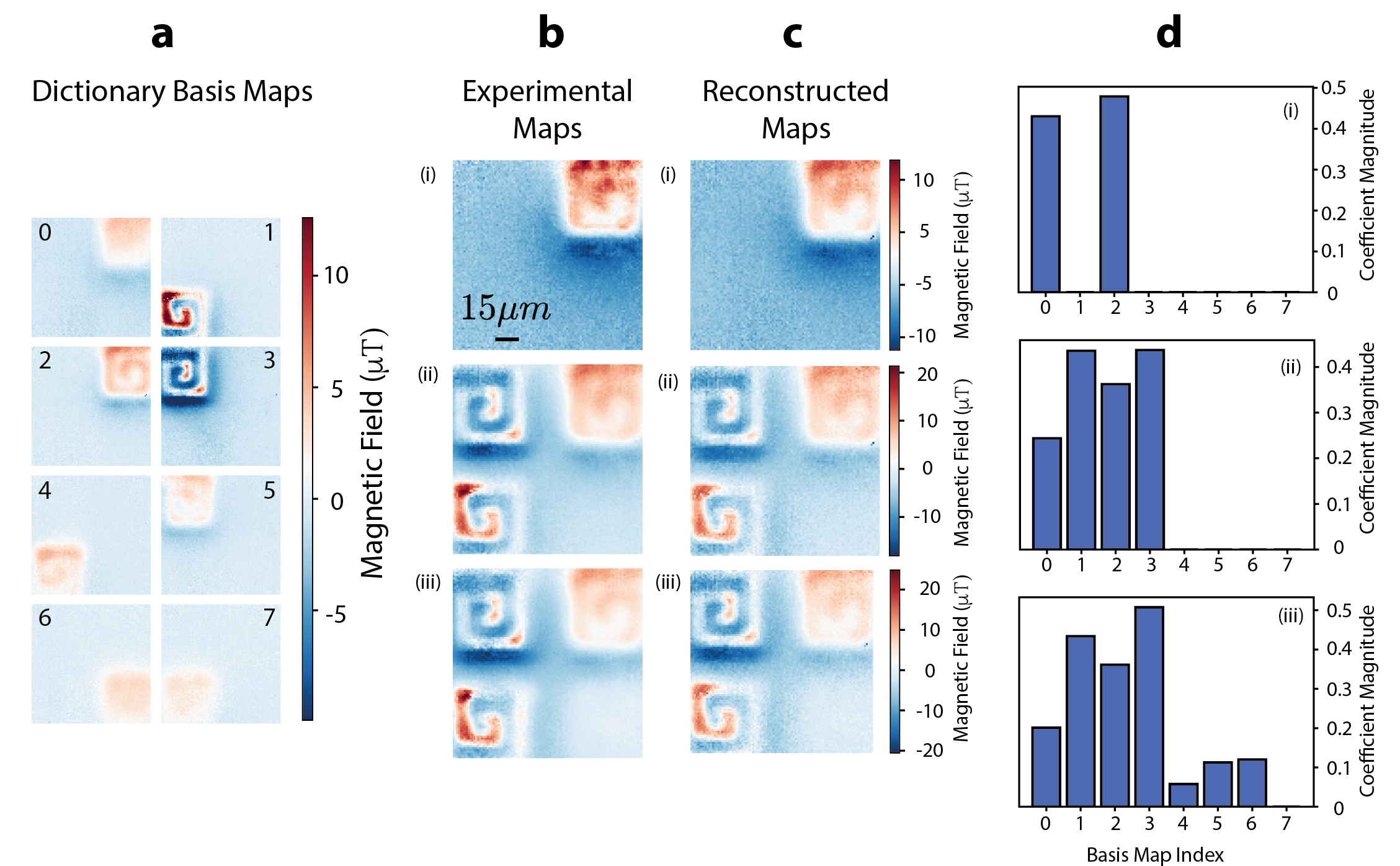}
	\caption{\textbf{Static Magnetic Field Reconstruction of Multilayer micro-coils Using LASSO Optimization}
		(\textbf{a}) Basis magnetic field maps of the eight individual micro-coils, forming the dictionary for reconstruction (\textbf{b}). Experimentally acquired magnetic field maps for different active coil combinations: (i) 2 active coils, (ii) 4 active coils, and (iii) 7 active coils.(\textbf{c}) Corresponding reconstructed magnetic field maps using LASSO optimization for (i) 2, (ii) 4, and (iii) 7 active coils.(\textbf{d}) Reconstructed sparse coefficients for each active coil in the three cases (i)–(iii), highlighting the relative contributions of each coil in the reconstruction process.}
	\label{fig2} 
\end{figure}

Most previous NV-based studies have focused on 2D current reconstruction due to the inherent non-uniqueness in mapping magnetic fields to current sources \cite{roth1989using}. Extending NV microscopy to focused 3D imaging applications across various domains has been relatively sparse. A fundamental challenge in reconstructing 3D magnetic field sources lies in the exponential spatial blurring introduced by the intrinsic nature of Biot-Savart’s law \cite{roth1989using}.
Moreover, the high correlation between magnetic field images from different micro-coil layers leads to ill-conditioned inversion problems, limiting the effectiveness of conventional linear reconstruction techniques such as matching pursuit \cite{AxonPotentialCommPhys2020}.\\
To accurately localize the current sources in three-dimensions, we employed LASSO-based sparse regression, leveraging a predefined dictionary of basis maps obtained from individual micro-coil field measurements. This method effectively deconvolves overlapping magnetic field contributions and isolates individual current sources with high accuracy.
Figure ~\ref{fig2}(a) illustrates the dictionary of individual micro-coil magnetic field maps, which serves as the basis for our reconstruction approach. Using this dictionary, we performed magnetic field reconstruction for any arbitrary combination of active coils by solving a sparse regression problem with LASSO. Experimental magnetic field images of micro-coil combinations, such as two, four, and seven active coils, are shown in Figure ~\ref{fig2}(b)(i-iii). The corresponding reconstructed maps are shown in Figure ~\ref{fig2}(c)(i-iii), demonstrating strong agreement with the experimental data. Additionally, Figure ~\ref{fig2}(d)(i-iii) presents the reconstructed coefficients for the active coils in each case, highlighting the ability of our method to accurately identify and separate the contribution of each individual micro-coil.\\
To quantitatively assess reconstruction performance, we evaluated the structural similarity index (SSIM)\cite{wang2009mean}, relative root mean square error (RRMSE), and Pearson correlation coefficient between the experimental and reconstructed maps. We observed high correlation across all cases $(0.95–0.98)$, along with strong SSIM values $(0.57–0.82)$ and moderate RRMSEs $(0.20–0.29)$, as shown in Table~\ref{Table1}.
\begin{table} 
	\centering
	\caption{\textbf{Static Reconstruction Performance Matrix}}
	\label{Table1} 
    \begin{tabular}{lccc} 
		\\
		\hline
		 & SSIM & RRMSE & Correlation Coeff\\
		\hline
		(i) & 0.5729 & 0.2734 & 0.9639\\
		(ii) & 0.8201 & 0.2051 & 0.9765\\
		(iii) & 0.7059 & 0.2897 & 0.9517\\
		\hline
	\end{tabular}
\end{table}

While overall reconstruction fidelity remains high, it is noteworthy that the 2-coil case (SSIM: 0.57) shows lower reconstruction fidelity compared to the 4-coils and 7-coils cases (SSIM: 0.82 and 0.70, respectively). This can be attributed to the greater standoff of the two active coils from the NV layer, both located in the second and third (deeper) layers. Increased coil-to-sensor distance causes stronger spatial blurring of magnetic fields, reducing structural similarity despite accurate coefficient reconstruction. A detailed analysis of this effect is provided in the Supplementary Sections 3 and 4.\\
Importantly, these metrics demonstrate that our LASSO-based approach maintains robustness and accuracy even as the number of active coils increases, highlighting its scalability to more complex multilayer configurations. By accurately reconstructing current sources from superimposed magnetic fields, this framework establishes a generalizable and computationally efficient method for 3D magnetic imaging.
To further evaluate the robustness of this approach under varying levels of complexity, we conducted an extended reconstruction study across 50 additional Experimental cases involving 2 to 7 active coils with varying layer distributions. Detailed statistical analysis, reconstruction accuracy, and regularization tuning are presented in Supplementary Section 3, providing further validation of the scalability and limitations of our reconstruction framework.

\begin{figure}[hbt!] 
	\centering
	\includegraphics[width=\textwidth]{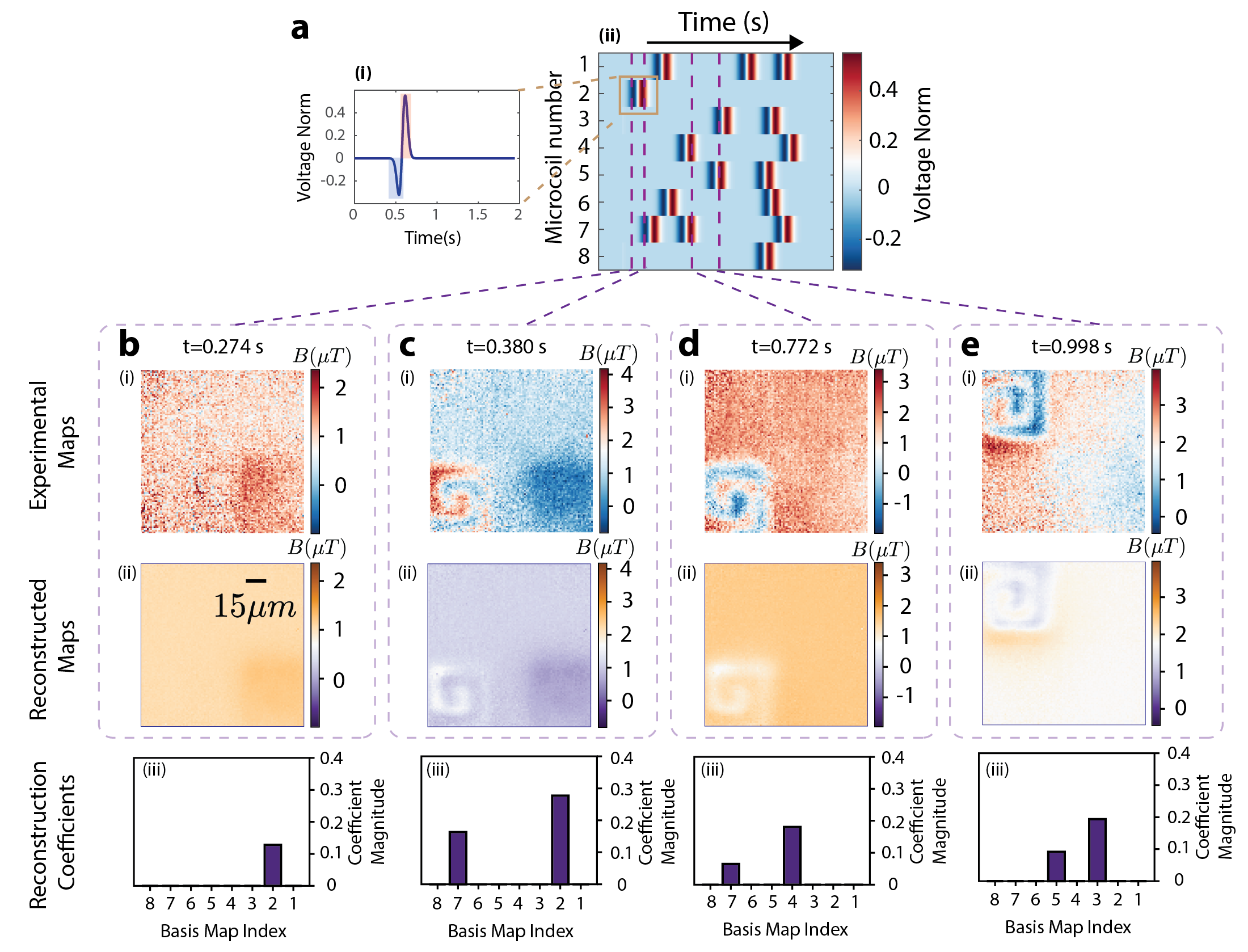}
	\caption{\textbf{Temporal reconstruction for High Sparsity case}
	(\textbf{a})(i) Current drive profile used to emulate action potential-like (AP-like) events with an $800 \mu A$ current driving the individual micro-coils. (ii) Waveform matrix across the eight micro-coils.(\textbf{b-e}) (i) Experimental magnetic field frames at the selected time points. (ii) LASSO-reconstructed magnetic field maps at the corresponding time points, demonstrating strong agreement with the experimental data. (iii) Reconstructed coefficients for the active micro-coils at each time point, highlighting the ability to track individual micro-coil contributions. The reconstruction accurately captures the temporal magnetic field variations, confirming the effectiveness of the LASSO-based sparse reconstruction approach.}
	\label{fig3} 
\end{figure}

\begin{figure}[hbt!] 
	\centering
	\includegraphics[width=\textwidth]{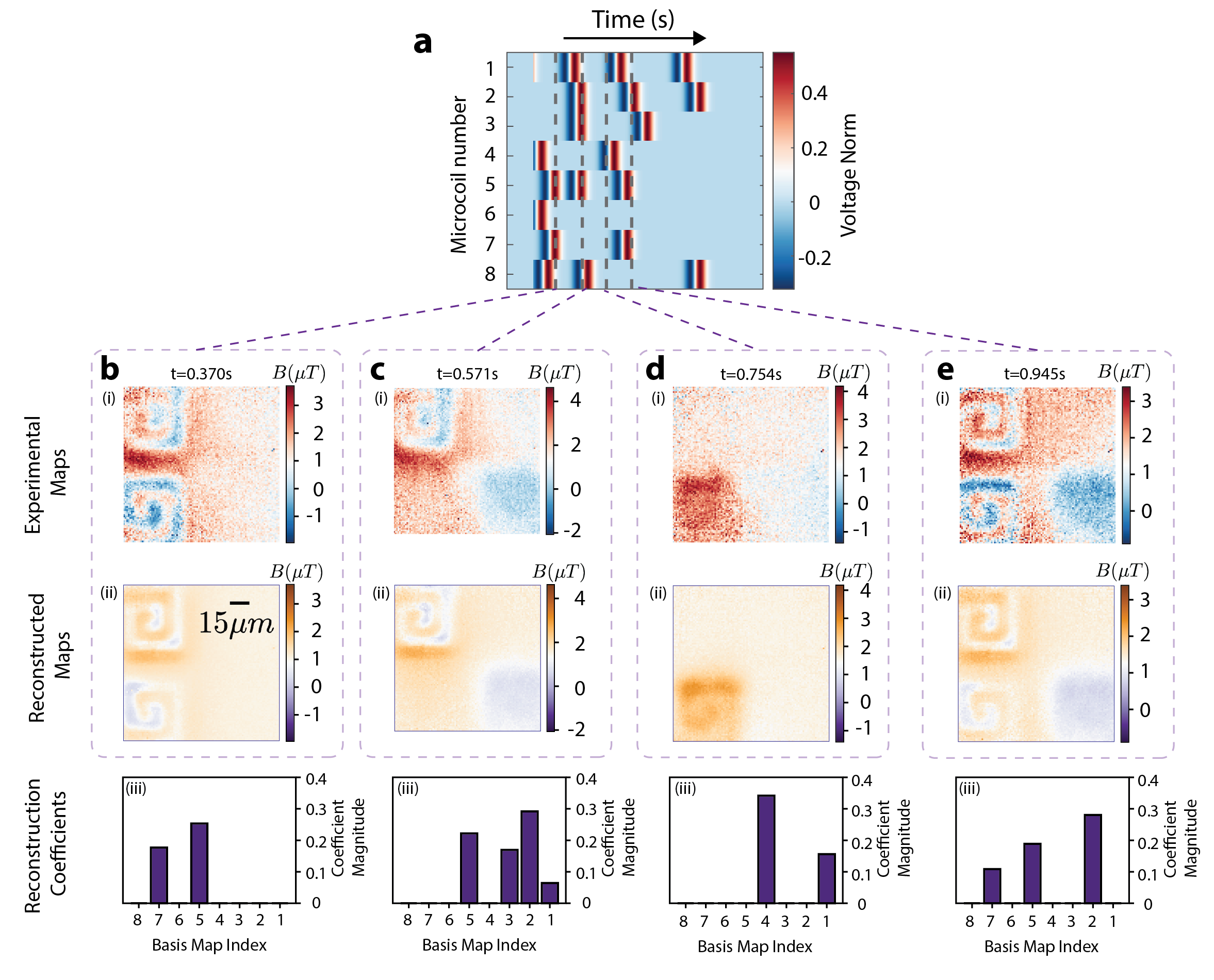}
	\caption{\textbf{Temporal reconstruction for Low Sparsity case}
		(\textbf{a})Current drive profile for the high firing rate scenario, showing the $800 \mu A$A current waveform matrix for the eight micro-coils.(\textbf{b-e}) (i) Experimental magnetic field frames for four selected time points, demonstrating the dynamic magnetic field variations. (ii) LASSO-reconstructed magnetic field maps corresponding to the experimental frames, with optimized regularization parameters ($\lambda$ = 0.2–0.25). The reconstructed maps align well with the experimental data. (iii) Reconstructed coefficients for the active micro-coils at each time point, illustrating the ability of the method to resolve contributions from individual micro-coils in real time.}
	\label{fig4}
\end{figure}

\paragraph{Temporal Imaging and Reconstruction}\mbox{}\\
Building upon the static reconstruction framework, we extended our methodology to handle temporally evolving magnetic field data acquired from the same multilayer micro-coil array. Using a predefined static dictionary of individual micro-coil magnetic field maps, we applied LASSO-based sparse regression to reconstruct current source distributions from selected temporal frames of dynamic magnetic field recordings.\\
To emulate neural-like dynamics, we implemented a synchronized temporal stimulus consisting of an action potential-like (AP-like) waveform [($ 300$ ms duration,$\ 800 \,\mu\text{A}$ peak current; Figure 3(a)(i))] applied simultaneously to all eight functional micro-coils. A refractory period was incorporated into the waveform design to prevent repetitive firing of the same micro-coil within a short interval. Two test cases were considered: a high-sparsity waveform matrix [Figure ~\ref{fig3}(a)(ii)], where only a few micro-coils are active at each time point, and a low-sparsity matrix [Figure \ref{fig4}(a)], involving greater overlap in spatiotemporal micro-coil activations. For both scenarios, four representative time frames were selected during periods of micro-coil activity, and frame-wise reconstruction was performed using the static dictionary  [Figure \ref{fig4}(b-e)].\\
Despite overlapping activations, distinct magnetic features from micro-coils in different layers remained resolvable, even in single-shot acquisitions. In NV-based magnetic imaging, temporal resolution is inherently coupled to per-pixel magnetic field sensitivity. We acquired millisecond-scale magnetic field images at a frame rate of 208 frames per second, balancing high temporal resolution with sufficient per pixel sensitivity of $\sim400~\mathrm{nT}/\sqrt{\mathrm{Hz}}$. Future improvements in diamond material quality and NV sensor optimization are expected to further enhance frame rates and magnetic sensitivity.\\
Reconstructing the underlying current distributions from these temporally evolving field maps poses greater challenges than in static imaging. Overlapping activations across layers, rapid current fluctuations, and potential increases in noise introduce ambiguity not present in static conditions. To address this, we applied LASSO-based sparse reconstruction to each frame, leveraging the sparsity of current sources and the static dictionary to infer micro-coil activations in real time. High-speed magnetic field videos (see Supplementary Movies S1 and S2) provide a well-controlled testbed to validate the reconstruction framework and demonstrate the feasibility of tracking millisecond-scale magnetic dynamics with high spatial and temporal resolution.\\
We used the same static basis maps (normalized to $\ 800 \,\mu\text{A}$) for all reconstructions, without introducing time-resolved dictionaries. For the high-sparsity case (Figure ~\ref{fig3}), LASSO was applied with optimized regularization parameters ($\lambda = 0.3–0.4$), yielding current distributions that closely resemble experimental magnetic field frames and effectively track individual coil activations. Quantitatively, the reconstructions yield structural similarity index (SSIM) values of 0.1615–0.2177, relative root mean square error (RRMSE) between 0.3604 and 0.5389, and Pearson correlation coefficients ranging from 0.2851 to 0.7605—indicating reasonable spatial and temporal fidelity. For the low-sparsity case (Figure ~\ref{fig4}), reconstruction remained robust with $\lambda = 0.2–0.25$, despite increased overlap among sources. These results demonstrate that the LASSO-based approach generalizes well across varying degrees of spatiotemporal sparsity.\\
Lower SSIM, higher RRMSE, and moderate correlation values in temporal reconstructions (compared to static cases) primarily arise from the use of a static dictionary. This dictionary, constructed from steady-state magnetic field maps of individual micro-coils, does not account for transient variations that may occur during rapid current switching or simultaneous activations—highlighting a fundamental trade-off between reconstruction speed and temporal precision.
Additionally, reconstruction fidelity tends to degrade when active micro-coils are located in deeper layers (second or third), due to their increased distance from the NV sensor plane. The larger standoff reduces spatial resolution and magnetic contrast, thereby lowering structural similarity. In contrast, activations in the topmost layer (closest to the sensor) yield sharper magnetic features and improved quantitative performance. A comprehensive analysis of reconstruction performance across different layers and sparsity conditions is provided in the Supplementary Sections 4.\\
Overall, this framework enables accurate, high-speed, frame-by-frame reconstruction of dynamic current activity in complex multilayer micro-coil systems. Our results underscore the effectiveness and flexibility of sparse recovery techniques under realistic spatiotemporal conditions, laying the groundwork for real-time current imaging applications.

\subsection*{Outlook}
In summary, we present a significant advancement in imaging dynamic 3-D magnetic  field distributions by combining high-resolution NV center-based magnetic microscopy with a LASSO-based reconstruction algorithm. Our approach enables spatio-temporal 3D localization of fluctuating current sources—an essential combination for investigating fast, localized phenomena such as neuronal signaling and transient faults in electronic systems. Unlike conventional methods such as MAGIC-UNet, which are primarily limited to 2D planar reconstructions,
our work demonstrates true 3D dynamic reconstruction using a biomimetic micro-fabricated circuit. At the core of our method is the use of LASSO optimization, which employs L1 regularization to promote sparsity. This makes it particularly effective for resolving current distributions where only a subset of potential sources is active at a given time. By selecting the most relevant basis maps from a dictionary of experimentally acquired magnetic field responses, the algorithm effectively de-convolves overlapping contributions from multiple layers. This approach provides distinct advantages over conventional methods. The Fourier method, for instance, is fundamentally unsuitable for reconstructing non-unique 3D magnetic field maps, especially when those maps are degraded by noise and increased standoff distances. Similarly, Tikhonov regularization tends to excessively smooth the results, compromising the resolution of sparse and localized features.

We further bench-marked our reconstruction framework against the pseudo-inverse method using finite element method (FEM)-simulated data from a large set of multilayer micro-coil arrays and 3D bent microwire geometries (See Supplementary Section 4). In both configurations, LASSO consistently outperformed the pseudo-inverse approach in accurate and spur-free isolation of active sources,
particularly under conditions involving high inter-source magnetic field overlap or increased sensor standoff. While the pseudo-inverse method yielded acceptable reconstructions in idealized scenarios with low noise and well-conditioned basis sets, it exhibited pronounced sensitivity to basis field coherence and experimental noise, often resulting in erroneous activation of inactive regions. In contrast, the sparsity-promoting nature of LASSO proved more robust in reconstructing sparse current distributions with minimal cross-talk, especially for complex 3D geometries.

Building on our earlier demonstration of 2D temporal imaging, we extend lock-in camera–based NV microscopy to full 3D imaging in this work, enabling millisecond-scale sensitivity to transient magnetic field dynamics. This advancement not only supports real-time tracking of dynamic current sources but also significantly accelerates the acquisition of experimental dictionary maps for reconstruction. Despite these capabilities, several key challenges remain. A primary limitation is the depth-dependent degradation in spatial resolution and signal-to-noise ratio, which affects reconstruction fidelity for deeper layers (See Supplementary Section 3). Moreover, strong coherence between basis fields from different micro-coil layers results in an ill-conditioned inverse problem, leading to potential leakage across inactive coils and difficulty in isolating true sources. Temporal fidelity can also be impacted when using a static dictionary, particularly under conditions of overlapping activations or rapid fluctuations. While we experimentally demonstrate high-fidelity spatio-temporal reconstruction using a complete dictionary in a controlled test case, further work is needed to fully realize the potential of this approach in a biological system. Specifically, it is essential to develop a robust noise model tailored to lock-in based magnetic field imaging, design effective dictionaries for scenarios where a complete dictionary is impractical, and enhance the overall sensitivity of the technique to enable reliable detection of weak bio-signals such as neuronal action potentials.

These challenges present clear directions for future improvement. Enhancing depth resolution and robustness may involve the use of more sophisticated or multi-height dictionaries, additional regularization techniques, or improved sensor design. Dynamic or adaptive dictionaries tailored to evolving current patterns could improve temporal reconstruction accuracy. Incorporating advanced noise suppression strategies, particularly those accounting for spatially correlated experimental noise, could further increase reconstruction fidelity. Scaling the approach to more complex geometries, such as densely packed 3D structures in integrated circuits or biological tissues, will require algorithmic innovation and further validation. Additionally, improving computational efficiency through optimized algorithms or hardware acceleration will be crucial for real-time, large-scale 3D reconstructions. Combining the sparsity-promoting power of LASSO with complementary machine learning techniques for field estimation and denoising may yield further performance gains.
In conclusion, the implications of our work are wide-ranging. Our method opens the door to label-free, high-resolution imaging of neural and cardiac activity, offering a new window into dynamic physiological processes. In semiconductor diagnostics, the ability to perform non-invasive, real-time 3D current mapping in multilayer integrated circuits provides a valuable tool for quality control, failure analysis, and potentially security assessment. More broadly, our framework offers a novel way to study charge transport in complex electronic systems and perform functional imaging in devices where existing 2.5D techniques fall short. By overcoming current limitations, leveraging its strengths, and expanding to additional image reconstruction modalities, this magnetic field reconstruction platform holds strong potential to evolve into a transformative technology for applications in both bio-sensing and microelectronics.

\subsection*{Methods}

\subsubsection*{Experimental Apparatus} \mbox{}\\
The experimental setup is described in Figure 1(a) illustrates the configuration of the NV-based widefield diamond microscope used for magnetic field imaging. The experiment utilizes a $4 \times 4 \, \text{mm}^2$ ultrapure diamond sample sourced from Element Six. The diamond is electronic-grade, containing 99.99\% $^{12}\text{C}$ with a $\sim1 \, \mu\text{m}$-thick layer doped with $^{15}\text{N}$ and an NV$^-$ concentration of approximately 1–2 ppm. The diamond crystal has a $\{100\}$ front facet orientation and a $\{110\}$ edge orientation. The NV$^-$ doped layer side of the diamond is mounted using Norland UV-curing adhesive at the center of the sample, which is wire-bonded to a custom PCB. A microwave loop antenna with a 3 mm diameter is placed on the opposite side of the diamond sample, enabling microwave excitation with frequencies sweeping around 2.87 GHz.

A 532 nm green laser, operating at 0.5 W, excites the NV$^-$ layer through a $100\times$ air objective with a numerical aperture of 0.9. The NV$^-$ centers emit red photoluminescence (PL), which is collected through the same objective, filtered using a notch filter (SEMROCK NF03-532E-25), and focused onto a wide-field lock-in camera (Heliotis Helicam C3). The lock-in camera enhances per-pixel sensitivity by filtering out noise outside the modulation frequency range and effectively utilizing the CMOS sensor's full-well capacity, thereby improving the dynamic range of the captured fluorescence. Each camera frame is synchronized with the microwave modulation through pulse sequences generated by a high-speed TTL pulse generator card (SpinCore PulseBlaster ESR-PRO 500 MHz). The system's magnification is $34\times$, with an overall field of view (FOV) of approximately $150 \, \mu\text{m} \times 150 \, \mu\text{m}$. Samarium-Cobalt (Sm-Co) ring magnets were used to provide a bias magnetic field.

\subsubsection*{Device Fabrication} \mbox{}\\
A novel sample emulating the spiking activity in mammalian neuron tissues was designed using small micro-coils, each with a size of $40 \, \mu\text{m}$ (with a $2 \, \mu\text{m}$ track width and $4 \, \mu\text{m}$ spacing), distributed across three layers on a Si-SiO$_2$ substrate. Each layer contains four micro-coils placed $20 \, \mu\text{m}$ apart in a square grid, covering an area of $110 \times 110 \, \mu\text{m}^2$. The resultant magnetic field at the desired standoff, corresponding to the three layers with different micro-coil combinations, was simulated using the magnetic field module in COMSOL Multiphysics.
The multilayer structure was realized using a combination of electron beam lithography, SU-8 spacer layers, and metal deposition on a thermally oxidized Si wafer. A detailed process flow is shown in Supplementary Figure S1
After fabrication, the individual micro-coils' DC resistance was measured using Proxima (Fast IV Measurement/ B1500). Each micro-coil on the Si/SiO$_2$ surface had a DC resistance of $\sim 1.1 \, \text{k}\Omega$, while the micro-coils on the SU-8 spacer layer had a DC resistance of $\sim 1.6 \, \text{k}\Omega$. The sample was then mounted with double-sided tape and wire-bonded to the metal bond pads on a custom-designed PCB.

\subsubsection*{Reconstruction Algorithm Using Pseudo-Inverse and LASSO}\mbox{}\\
We consider a linear model for reconstructing magnetic field maps using a set of basis maps:
\[
\mathbf{y} = \mathbf{A} \mathbf{x} + \boldsymbol{\epsilon}
\]
where:
\begin{itemize}
    \item $\mathbf{y} \in \mathbb{R}^{M}$ represent the Experimental test map/ simulated map (flattened into a vector).
    \item $\mathbf{A} \in \mathbb{R}^{M \times N}$  dictionary matrix, where each column is a flattened basis map.
    \item $\mathbf{x} \in \mathbb{R}^{N}$ is the coefficient vector that represents the contribution of each basis map.
    \item  ${\epsilon}$ is the experimental noise
\end{itemize}
\paragraph{Pseudo-Inverse Reconstruction}\mbox{}\\
This approach finds the least-squares solution to the linear system using the Moore–Penrose pseudo-inverse:
\[
\mathbf{x} = \mathbf{A}^{+} \mathbf{y}
\]
Here $\mathbf{A}^{+} = (\mathbf{A}^{T}\mathbf{A})^{-1}\mathbf{A}^{T} $
if (\boldmath$A$ has full coloumn Rank)\\
This solution minimizes the Euclidean norm
\[
\underset{\mathbf{x}}{\text{minimize}} \quad  \|\mathbf{A x} - \mathbf{y}\|_2^2
\]
\subsubsection*{LASSO Reconstruction}\mbox{}\\
LASSO (Least Absolute Shrinkage and Selection Operator) is a regression method that promotes sparsity in the solution by applying L1 regularization. It is used to estimate the contribution of individual basis maps to a measured test map while enforcing sparsity to ensure that only the most relevant basis maps are selected.\\

The LASSO optimization problem is then given by:
\[
\underset{\mathbf{x}}{\text{minimize}} \quad \frac{1}{2} \|\mathbf{A x} - \mathbf{y}\|_2^2 + \lambda \|\mathbf{x}\|_1
\]
where ${\lambda}$ is the regularization parameter controlling the sparsity of x.\\
\subsubsection*{Static Field Reconstruction}\mbox{}\\
Both the pseudo-inverse and LASSO approaches follow the same preprocessing steps:
\begin{enumerate}
    \item The basis maps are loaded from experimental datasets, where each basis map corresponds to the field contribution of an individual micro-coil.
    \item The test map is computed as the difference between the magnetic field maps in the I-on and I-off states for a given coil combination.
    \item Both the basis maps and test maps are mean-centered and standardized to have unit variance to ensure numerical stability.
    \item For pseudo-inverse: coefficients are estimated using the pseudo-inverse of the basis matrix.
    \item For LASSO: The regression model is initialized with a chosen value \boldmath$\lambda$ (typically tuned) and the model is trained to find sparse coefficients \boldmath$x$ by solving the optimization problem.
    \item The estimated coefficients are used to linearly combine the basis maps, forming the reconstructed test map.
    \item The reconstruction is then rescaled back to the original scale of the experimental data.
\end{enumerate}
\subsubsection*{Temporal Field Reconstruction}\mbox{}\\
The temporal reconstruction follows the same steps as the static case but is applied to individual frames at different time points:
\begin{enumerate}
    \item Experimental time-resolved data from 300 frames is loaded, using the same dictionary of basis maps for all time points.
    \item Each frame at a specific time point is treated as an independent test map \boldmath$y_t$ and baseline correction is applied by subtracting the mean field from frames with no current waveform applied.
    \item Normalization and saturation correction are performed to mitigate intensity variations and oversaturated pixels.
    \item Basis and test maps are mean-centered and standardized for numerical stability.
    \item LASSO regression is applied with fine-tuning of the regularization parameter to balance sparsity and stability.
    \item The time-dependent coefficients, \boldmath$x_t$, are used to linearly combine basis maps, reconstructing the test map for each time point.
    \item The reconstructed map is then rescaled back to the original scale of the experimental data.
\end{enumerate}

\section*{Acknowledgments} The authors are grateful to D. Shishir, Prof. Udayan Ganguly, Prof. Pradeep Sarin, Prof Kantimay Dasgupta and Prof. Swaroop Ganguly for their valuable insights on experimental methodology and device fabrication.

\paragraph*{Funding}
K.S. acknowledges funding from DST National Quantum Mission, TCS Research,  AOARD grant number FA2386-23-1-4012, DST-SERB Power Grant SPG/2023/000063, and DST Quest DST/ICPS/QuST/Theme-2/2019/Q-58. A.B. and K.S. acknowledge support from IITB Nanofabrication facility. A.B. acknowledges funding from I-Hub Quantum Technology Foundation's Chanakya Doctoral Fellowship.

\paragraph*{Author contributions:}

KS, AB, and MP conceived the idea. AB was responsible for device fabrication, conducting the experiments, and performing the reconstruction. MP provided initial support with experimental data collection and reconstruction simulations. MM provided the diamond NV layer samples. AR supervised the source localization and reconstruction efforts. AB drafted the manuscript with input from all authors. KS supervised the overall project.

\paragraph*{Competing interests:}
There are no competing interests to declare.

\paragraph*{Data and materials availability:}
The data that support the findings of this study are available from the corresponding author upon reasonable request.


\end{document}